\documentstyle[epsf,PASJadd]{PASJ95} 
\newcommand{\vol}{49}
\newcommand{\no}{6}
\newcommand{\lett}{}
\newcommand{\spage}{}
\newcommand{\rdate}{1996 08 28}
\newcommand{\adate}{1997 09 15}

\def\tatehaba2{\rule[-1.5mm]{3mm}{5mm}}
\def\tatehaba3{\rule[1mm]{2mm}{0.5mm}}

\onecolumn

\begin{document}

\title{Wide-Band X-Ray Spectra and Images of the Starburst Galaxy M82}

\author{Takeshi Go {\sc Tsuru},$^1$ 
Hisamitsu {\sc Awaki},$^1$ Katsuji {\sc Koyama},$^{1,3}$ 
and Andy {\sc Ptak$^2$} \\
{\it $^1$ Department of Physics, Faculty of Science, Kyoto University, 
Sakyo-ku, Kyoto 606-01} \\
{\it E-mail (TGT): tsuru@cr.scphys.kyoto-u.ac.jp} \\
{\it $^2$ NASA/GSFC, Code 662, Greenbelt, MD 20771, USA} \\
{\it $^3$ CREST: Japan Science and Technology Corporation (JST), 
4-1-8 Honmachi, Kawaguchi, Saitama 332}}
\abst{The ASCA results of the starburst galaxy M82 are presented. 
The X-rays in the 0.5--10 keV band exhibit a thin thermal spectrum 
with emission lines from highly ionized magnesium, silicon, and sulfur,  
as well as a hard tail extending to higher than 10keV energy. 
The soft X-rays are spatially extended, while the hard X-rays show 
an unresolved point-like structure with possible a long-term flux variability. 
The flux ratio of the emission lines and the spatially extended structure 
in the low-energy band indicate 
that at least two-temperature thin thermal plasmas are present. 
The abundances of the oxygen, neon, magnesium, silicon, sulfur, 
and iron in the thin thermal plasmas are found to be 
significantly lower than the cosmic value. 
Neither type-Ia nor type-II supernova explosions can 
reproduce the observed abundance ratio. 
The origin of the unresolved hard X-rays is uncertain, 
but is probably an obscured low-luminosity AGN.}

\kword{Galaxies: individual(M82) --- Galaxies: abundance 
	--- Galaxies: active --- Galaxies: X-rays --- X-rays: spectra}

\maketitle
\thispagestyle{headings}

\section{Introduction}
Starburst galaxies comprise a sub-class of young galaxies, 
which exhibit violent star formation activities, 
mainly near to the center of the galaxy. 
The starburst activity may produce hot plasmas, 
due either to frequent supernova explosions 
or to strong stellar winds from young massive stars. 
The hot plasmas can partly escape to intergalactic space as the galactic wind. 
Thus, the starburst galaxies would be good laboratories 
to study a possible interplay between star formation, 
heating of interstellar and intergalactic gas and 
chemical evolution over a wide range of the astronomical scales; 
from supernova and star clusters to the inter-galactic scale. 
X-ray observations should provide a key to study these issues, 
because the hot plasmas preferentially emit their energy in the X-ray band, 
giving information about the plasma temperature and density. 
Also, X-ray spectroscopy provides a direct method 
to determine the chemical composition.  

M82 has been known to be the most active starburst galaxy near the galaxy, 
and, together with its proximity of 3.25~Mpc distance, 
is placed to be the best-studied object (e.g., Rieke 1980). 
An X-ray image of M82 was firstly provided with Einstein HRI observation 
(Watson et al. 1984). 
They found, in addition to several unresolved sources, 
diffuse X-rays extending along the minor axis of the M82 galaxy. 
From the IPC data, Fabbiano (1988) determined 
the temperature of the extended emission to be 1.2--2.7 keV, 
while the MPC data revealed the existence of a high-temperature component, 
possibly associated with or near to the galactic nuclei. 
Although these authors failed to distinguish 
whether the diffuse emission is thermal or non-thermal, 
they interpreted that the X-rays are due to a galactic wind 
from the center of M82 at a rate of $\sim 0.7$~{\it M$_{\odot}$}~yr$^{-1}$. 
More detailed studies were made with ROSAT PSPC and HRI observations 
(Watson et al. 1994; Bregman {\rm et al.}\  1995). 
ROSAT HRI detected a point source at the galactic nucleus 
with {\it L$_{\rm X}$}$=10^{40}$~{\rm erg~s$^{-1}$}. 
It even exhibited a flux variability on a time scale of about two weeks 
(Collura {\rm et al.}\  1994). 
Strickland {\rm et al.}\  (1997) extensively studied with ROSAT PSPC and HRI, 
and found that the diffuse emission is really thin thermal emission 
with a temperature of {\it k$_{\rm B}T$}$=0.6$~keV\ at the center, 
and decreasing to {\it k$_{\rm B}T$}$=0.4$~keV 
along the minor axis with absorption of 
{\it N$_{\rm H}$}=$1\times 10^{21}$~{\rm cm$^{-2}$}. 
They also reported that the nuclear region exhibits 
two temperature components of {\it k$_{\rm B}T$}$=0.76$~keV 
with {\it N$_{\rm H}$}$=0.9\times 10^{21}$~{\rm cm$^{-2}$} 
and {\it k$_{\rm B}T$}$>6.2$~keV 
with {\it N$_{\rm H}$}$=6\times 10^{21}$~{\rm cm$^{-2}$}; 
the absorption of the hotter component 
is larger than those of the diffuse emission. 
Ginga found that the X-ray spectrum in the 2--10~keV band 
could not be fit with a power-law model, 
but could be nicely fit with thermal bremsstrahlung at a temperature of 
{\it k$_{\rm B}T$} $=5.75^{+0.58}_{-0.57}$keV absorbed by a cold column of 
{\it N$_{\rm H}$}$=8.9^{+2.6}_{-3.0}\times 10^{21}$~{\rm cm$^{-2}$}\ 
(Tsuru 1992). 
The author also reported a marginal detection of the iron K emission line. 
However, the center energy was not determined well; 
we were thus unable to distinguish 
whether the hard X-rays are of thermal or non-thermal origin. 
All of these previous observations have been limited 
in either imaging capability (for the hard X-rays) or energy resolution. 
The first experiment with moderately high-energy resolution 
in the hard X-ray band was made with BBXRT. 
Petre (1993) reported, although only preliminary, 
the first detection of K-shell transition lines of 
highly ionized magnesium, silicon, but no iron. 
However, the limited statistics prevent the author 
to obtain any detailed information.
ASCA is the first X-ray satellite having wide energy-band imaging 
and spectroscopy capability with reasonably high statistics, 
and hence may be a suitable instrument 
for studying the nature of X-rays from M82. 
Since the image quality of the low-energy band is worse than 
those of the previous satellites, 
we put our emphasize more on the wide-band spectra than on the imaging aspects.
Some preliminary reports concerning the ASCA observation 
have already been made by Tsuru {\rm et al.}\  (1994), 
Awaki {\rm et al.}\  (1996), Tsuru {\rm et al.}\  (1996), 
and Ptak {\rm et al.}\  (1997).
Ptak etal (1997) focused on the SIS data and 
Moran, Lehnert (1997) reported results mainly from GIS and PSPC. 
This paper reports on the results of a combined study 
using both of GIS and SIS. 

\section{Observation and Data Reduction}
M82 was observed with ASCA from 1993/04/19 22:22 to 04/20 07:58 
using two solid-state imaging spectrometers (SIS~0 and SIS~1) 
with the 4CCD Faint and Bright modes and 
two gas-imaging spectrometers (GIS~2 and GIS~3) with the normal PH mode; 
both are separately placed on each focal plane of 
four identical thin-foil X-ray mirrors (XRT). 
Details concerning the instruments are described 
in Burke {\rm et al.}\  (1991), 
Ohashi {\rm et al.}\  (1996), Makishima {\rm et al.}\  (1996), 
and Serlemitsos {\rm et al.}\  (1995), 
while a general description of ASCA can be found in 
Tanaka {\rm et al.}\  (1994). 
The data were screened with the standard selection criteria 
with XSELECT version 1.3 so as to exclude data 
affected by the South Atlantic Anomaly, 
earth occultation, and regions of low geomagnetic rigidity. 
We eliminated the contamination by the bright earth, 
removed hot and flickering pixels from  the SIS data 
and rejected particle events from the GIS data 
using the rise-time discrimination technique. 

After these screenings, 
we obtained effective accumulation times of 18.4 and 16.0~ks 
for SIS and GIS, respectively, out of the total exposure of 22.5~ks. 
The counting rates of SIS and GIS were 1.3 counts~s$^{-1}$~SIS$^{-1}$
and 0.86 counts~s$^{-1}$~GIS$^{-1}$, respectively. 


\section{Analysis and Results}
\subsection{Spectral Analysis}
We collected X-ray photons 
from the $6'$ radius circle around the center of M82 image. 
The background spectrum  of the SIS data was  taken 
from a square region of $17'.14\times 17'.14$, 
excluding the region $r>7'.65$ from the center of M82, 
while that of GIS was collected 
from the $7'.65<r<18'.0$ annulus around M82. 
After subtracting the background, 
we made a combined SIS spectrum 
(SIS~0 and SIS~1 were added, SIS~0+1 hereafter). 
The data from GIS~2 and GIS~3 were  also combined (GIS~2+3 hereafter). 
Figure~1 shows the thus obtained SIS~0+1 and GIS~2+3 spectra. 
We used the XSPEC version 8.50 or 9.00 for the following spectral fits. 

The number of atoms per hydrogen for the cosmic metal abundance 
adopted in this paper are 
$0.0977$, $3.63\times 10^{-4}$, $1.12\times 10^{-4}$, 
$8.51\times 10^{-4}$, $1.23\times 10^{-4}$, $3.80\times 10^{-5}$, 
$3.55\times 10^{-5}$, $1.62\times 10^{-5}$, $3.63\times 10^{-6}$, 
$2.29\times 10^{-6}$, $4.68\times 10^{-5}$, and $1.78\times 10^{-6}$ 
for He, C, N, O, Ne, Mg, Si, S, Ar, Ca, Fe, and Ni, respectively 
(Anders, Grevesse 1989). 

\subsubsection{Low energy emission lines}
From figure~1, we can see several resolved emission lines 
above 1.2 keV and unresolved complex lines near and below 1 keV.  
We have determined the center energies and fluxes of the resolved lines
by fitting with a model of narrow Gaussian lines on a power-law continuum 
in a limited energy band. 
The line energies and fluxes are shown in table~1 
along with the most probable atomic-line candidates. 
The presence of emission lines from 
highly ionized magnesium, silicon and sulfur 
indicates  that the X-rays are, at least partly, 
attributable to a thin thermal plasma. 

The line fluxes of H- and He-like atoms give the population of 
the ionization states, or the ionization temperature of the atom.
The ionization temperature is 
a function of the electron temperature {\it k$_{\rm B}T$} 
and an ionization parameter $n_{\rm e}t$, 
where $n_{\rm e}$ and $t$ are, respectively, 
the electron density and elapse time of the plasma 
with temperature of {\it k$_{\rm B}T$}.
We note that at ${\rm log}(n_{\rm e}t)>12.5$, 
the ionization temperature becomes nearly equal to the electron temperature 
(collisional ionization equilibrium or CIE in short), 
and that many astronomical plasmas are far from CIE; 
they are still in the ionizing phase, called non-equilibrium ionization, 
or NEI in short. 

From the silicon and magnesium lines listed in table~1, 
we made a two-dimensional parameter map of $n_{\rm e}t$ 
and the electron temperature {\it k$_{\rm B}T$}\ (figure~2) using the code of 
ionization in a plasma developed by Masai (1984, 1994). 

No overlapping region for silicon and magnesium was found, 
indicating that the plasma was not at a single temperature.
We thus proceed to the spectral analysis with multi-temperature components, 
either in CIE or NEI.

\subsubsection{Spectral fitting}
The SIS~0+1 and GIS~2+3 spectra were fitted simultaneously
by a 2-temperature Raymond Smith thin thermal plasma model of CIE 
(Raymond, Smith 1977) at the temperatures of 1.1 and 0.5 keV, 
which were estimated from the line flux ratios given in figure~2 
[see Si and Mg temperatures at ${\rm log}(n_{\rm e}t) > 12.5$]. 
We found a large excess in the data above the 3~keV bands.
This immediately indicates that the spectra require 
another hard component whose temperature is above 10~keV. 
We thus conducted a simultaneous fitting to the SIS~0+1 and GIS~2+3 spectra 
with a three-temperature model (3-Raymond Smith or 3-RS): 
the soft, medium and hard components, hereafter. 

As shown in the next section, the soft component has an apparent extension. 
On the other hand, the emitting region of medium 
and the hard components are compact at the center region 
of the soft component. 
We thus assume the following model spectrum: 
\begin{eqnarray}
\lefteqn{N_{\rm H}({\rm Whole})\times 
        [ 
	{\rm Thermal(Soft)} } \nonumber \\
	& + & N_{\rm H}({\rm Medium}) \times {\rm Thermal(Medium)} \nonumber \\
	& + & N_{\rm H}({\rm Hard}) \times {\rm Thermal(Hard)} 
	] .
\end{eqnarray} 

We fit the spectra with a 3-temperature  model 
using independent absorption columns for each component. 
Since the light metals, such as silicon and sulfur, are expected to be 
fully ionized at the high-temperature (hard) component, 
the abundances of Si and S in the hard component are not well constrainted. 
We therefore fixed the abundance ratios in the hard component 
to be solar ones. 
For the other components, the low- and medium-temperature plasmas, 
we let the metal abundances be free. 
With this model, the metal abundances of the medium and hard components
are reasonably constraint, but not for the low-temperature component. 
No additional absorption for the medium component to the whole absorption 
(also for the soft component) is required with  an upper limit of  
$3.2\times 10^{21}${\rm cm$^{-2}$}.
We therefore assume that the metal abundances between 
the soft and the medium components are the same 
with no additional absorption for the medium component.
This model gave a reasonable $\chi^2$/d.o.f. of $444.8/372$.
The best-fit results are given in figure~1, 3 and table~2. 
We note that the metal abundances of the medium components are very similar 
to those obtained using the previous model, 
in which the abundances for both the soft and medium components are free. 
The relative normalization of GIS~2+3 to SIS~0+1 is free, 
and turns out to be 1.09, 
which is within the error on the cross-calibration of GIS and SIS.
In short, we conclude, from a series of spectral fittings, 
that the abundances of the medium and hard components are well constrainted, 
and those of the soft component are consistent 
with being the same abundances of the medium. 
Additional absorption is only required for the hard component.


An alternative model, 
in which the hard component is replaced by a power-law model, 
gave a good fit with a $\chi^2$/d.o.f. of $454.2/373$. 
In this case also, additional absorption is required only 
for the hard power-law component. 
Thus, no essential difference was found between the non-thermal (power-law) 
and thermal (RS) models for the hard component; 
hence, whether the hard component is thermal or non-thermal remains uncertain.
The best-fit parameters are given in table~3. 

Moran, Lehnert (1997) also reported 
that the X-ray spectrum of M82 can be described by three components: 
two thin thermal plasmas and a power law dominating at the high-energy band. 
However, this result is different from ours; 
the large cold absorption column to the medium component, 
or the ``warm thermal component'' in Moran, Lehnert (1997), 
is not required in our fitting. 
This discrepancy would be attributable to the assumption 
on a metal abundance. 
Moran, Lehnert (1997) assumed solar abundances for the two thermal components. 
This model, however, does not fit the SIS line structures 
with higher energy resolution than those of GIS and PSPC. 

\subsubsection{Upper limit on iron K emission line}
Since the iron K-shell emission line provides key information 
concerning a plasma diagnostics of a temperature higher than a few keV, 
we made a spectral fitting with a thermal bremsstrahlung model plus 
a Gaussian line around 6--7~keV in a limited energy band of 4--8~keV. 
Since the energy band was selected to be reasonably narrow, 
the result did not depend on the spectral model of the continuum component. 
We found no significant emission line. 
The upper limit is shown in figure~4 
as two-dimensional (Line energy--equivalent width) contours. 
Two-dimensional contours determined with Ginga are also given (Tsuru 1992). 

\subsection{Imaging Analysis}
Figure~5 shows the SIS~0 images in the 0.5--0.8~keV, 
1.2--1.8~keV and 3--10~keV bands:  the energy bands which represent 
the soft, medium and hard components, determined 
from the 3-RS fitting (see figure 1). 

We found that the 0.5--0.8~keV band X-rays show significant spatial extension, 
particularly along the minor axis of the optical image of M82. 
The radial profile in the range of $r=0'.5$--$5'$ 
can not be represented by the point spread function (PSF in short) 
with $\chi^2$/d.o.f$=35.9/12$. 
This, together with the spectral information, 
indicates that the soft component is a diffuse thermal plasma 
extending larger than the optical image of M82. 

The radial profile of the 3--10~keV band, on the other hand, 
is consistent with the PSF ($\chi^2$/d.o.f$=8.8/12$). 
The upper limit on the spatial extension is $0'.5$ in radius. 
Thus, the hard component is point-like within the ASCA resolution. 

The radial profile of the 1.2--1.8~keV band image is not fitted with
the point spread function with $\chi^2$/d.o.f$=27.6/12$. 
Therefore, the medium component is extended, 
but its extension may be smaller than that of the soft component.
The peak positions of the 1.2--1.8~keV and 3--10~keV bands agree with 
each other within the spatial resolution. 
The 0.5--0.8~keV band image is more complex, 
as is shown in the closed view (figure~6). 
A couple of peaks around the galactic center are seen 
in the 0.5--0.8~keV image. 
However, neither peak agrees  with 
the peak of the 1.2--1.8~keV and 3--10~keV bands.

\section{Discussion}
We found that the X-rays of M82 have a very complex structure, 
comprising extended soft and medium-temperature 
plasmas and point-like hard X-rays near to the center of M82. 
Based on the ASCA results, and referring to previous X-ray observations, 
we try to give possible implications on the nature of M82 X-rays. 

\subsection{Hard X-Rays}
\subsubsection{Time variability}
The spatial extension of the hard X-rays near to the center of M82 is 
constrainted to be less than 0.5~kpc ($0'.5$) radius. 
Moran, Lehnert (1997) already reported that no short time variability 
was detected in the light curve of GIS~2+3 with an energy band of 3.5--10~keV. 
We also detected no short time variability in SIS~0+1. 

The hard component should be the same as those observed 
with previous non-imaging hard X-ray instruments, 
from the first detection of Uhuru to recent Ginga scanning observations. 
Since most of these previous observation might have suffered 
from possible contamination of M81, lying about 1 degree away from M82, 
we compare them with the results from relatively 
small field-of-view experiments of EXOSAT ME 
and scanning observation of Ginga (Schaaf {\rm et al.}\  1989; Tsuru 1992). 
Both of them safely avoided M81 contamination. 
When calculating the X-ray flux for each experiment, 
we adopted the best-fit spectral model to each observational spectrum. 
The results concerning the X-ray fluxes in the 2--10~keV band are, 
$(2.3 \pm 0.2)\times 10^{-11}$, $(3.4 \pm 0.1)\times 10^{-11}$, 
$1.9\times 10^{-11}$ {\rm erg~cm$^{-2}$~s$^{-1}$}\ 
for EXOSAT ME, Ginga and ASCA, respectively. 
Assuming that M81 contributed 15\% of the flux obtained from EXOSAT ME 
observation, the M82 flux observed with EXOSAT ME becomes 
$(2.0 \pm 0.2) \times 10^{-11}$ {\rm erg~cm$^{-2}$~s$^{-1}$}\ 
(Schaaf {\rm et al.}\  1989). 

We see a significant variation of about a factor of 2 
in the X-ray luminosities. 
Since 90\% of the flux in the 2--10~keV band derived from ASCA was 
emitted from the hard component, 
we conclude that the hard component of M82 has been variable 
with a long time scale of years. 
Adding to this, a significant spectral change from Ginga to ASCA is seen. 
The X-ray spectrum at the time of the Ginga observation 
was not fit with a power-law model, 
but was nicely fit with thermal bremsstrahlung at a temperature of 
{\it k$_{\rm B}T$} $=5.75^{+0.58}_{-0.57}$~keV, absorbed by a cold column of 
{\it N$_{\rm H}$}$=8.9^{+2.6}_{-3.0}\times 10^{21}$~{\rm cm$^{-2}$}\ 
(Tsuru 1992), which is different from the spectral shape obtained with ASCA. 

These two results concerning the flux and spectral shape indicate 
that the source of the hard component is a time variable. 
The time scale gives the upper limit on the size of the hard X-ray emitter 
to be less than 1~pc, which is a far more severe constraint 
than that obtained from imaging analysis.  

The hard component might be a summation of 
the unresolved galactic point sources. 
However, at least half of the flux in the 2--10~keV band 
observed with Ginga was time variable with a timescale of several years, 
which immediately indicates that at least half the time variable was 
due to a point source (Tsuru 1992). 

\subsubsection{Comparison with ROSAT observation}
The spectrum of the nuclear region obtained with ROSAT PSPC is 
reproduced by a two-thermal bremsstrahlung model of 
{\it k$_{\rm B}T$}$>6.2$~keV and {\it k$_{\rm B}T$}$=0.76$~keV 
(Strickland {\rm et al.}\  1997). 
This may be consistent with our interpretation: 
a hard X-ray source at the M82 center 
and diffuse soft and medium X-rays around the center. 

Near to the M82 center, ROSAT HRI found three point sources; 
among them, source~8 is the brightest, 
by a factor of 25, than any other point sources 
(Strickland {\rm et al.}\  1997). 
Therefore, this HRI source could be a possible candidate of 
the ASCA hard X-ray source. 
The position error of the ASCA image, 
however, is as large as about $0'.5$ 
and no reference X-ray star in the ASCA field-of-view is available. 
Furthermore, as we already noted, 
the peak of the hard and medium components does not 
agree with that of the soft component. 
Thus, identification of the hard X-ray source to source~8 is debatable. 
Nevertheless, we discuss the possibility of the hard component 
to be source~8.  
From the best-fit spectral model parameters given in table~2 
and HRI spectral response, 
the flux of the hard component obtained from our observation was converted to 
a HRI counting rate of $0.018$~{\rm counts~s$^{-1}$}. 
This value is 2--3 times smaller than the observed ROSAT HRI flux 
of source~8, $0.050$~{\rm counts~s$^{-1}$}, 
but within the variable range obtained during the HRI observation 
(Collura {\rm et al.}\  1994).

\subsubsection{Origin of the hard X-rays}
Rieke {\rm et al.}\  (1980), Schaaf {\rm et al.}\  (1989), 
and Moran, Lehnert (1997) 
argue that an inverse-Compton X-ray emission scattered from infrared photons 
is an origin of the hard X-rays. 
However, the size of the emitting region, 
evaluated from the observed long-term variability, 
is estimated to be only $\sim$ 1~pc, 
far smaller than the previous estimation by an order of two or three. 
Thus, the contribution from this process 
to the total X-ray emission could be ignored. 

The brightest ROSAT HRI source~8, 
can be identified with a bright radio source, 
either 41.5+597 or 41.95+575 (Strickland {\rm et al.}\  1997). 
The radio source 41.95+575 may be attractive 
as a counterpart of the time variable ASCA hard X-ray source 
or ROSAT HRI source~8, 
because its radio flux is variable with a decreasing rate of 9\% per a year. 
Adding to this, its synchrotron self-absorption frequency is also decreasing 
and the source shows a shell-like structure with a diameter of 0.3~pc 
(Muxlow {\rm et al.}\  1994). 
All of these results led Kronberg {\rm et al.}\  (1985) 
to propose that 41.95+575 is a type-II supernova, 
exploded in a very dense region with $n_0\sim 10^7$~{\rm cm$^{-3}$}. 
Terlevich (1994) also modeled 
that 41.95+575 is an SNR of the type-II supernova 
which occurred in $\sim 1955$. 
This putative  young SNR (41.95+575) may produce hard X-rays,
as is found in other young supernova remnants, such as SN~1993J in M81 
and SN~1978K in NGC~1313 (Kohmura {\rm et al.}\  1994; 
Petre {\rm et al.}\  1994).
The initial X-ray luminosity of theses SNRs are 
about $10^{40}$~{\rm erg~s$^{-1}$}\ (SN1993J), 
which is more or less comparable to the hard X-ray flux of M82.
However, unlike the above sample of young SNRs, 
the hard X-ray luminosity of M82 has decreased from Ginga to ASCA 
by factor of 2, while the temperature has increased from 5.75~keV to 14~keV. 
Furthermore, no strong iron-K line was found. 
Thus, the X-rays from a possible young SNR (the radio source 41.95+575) is 
unlikely for the origin of the hard component. 

One may argue that the hard source would be a super-Eddington source,
which is often found in some spiral galaxies. 
However, the luminosity of the hard component of 
$4 \times 10^{40}$~{\rm erg~s$^{-1}$} is far larger than 
that of any super-Eddington source 
($10^{39}$ to $10^{40}$~{\rm erg~s$^{-1}$}, 
see e.g. Okada {\rm et al.}\  1994).

Another possible counterpart would be 41.5+597, 
which is highly variable in the radio band; 
it was $\sim 7$~mJy in 1981 and decreased by $\sim 100$\% within a year 
(Kronberg, Sramek 1992). 
An HST observation showed that there is a very bright spot 
at the position 41.5+597 (region~E in O'Connell {\rm et al.}\  1995). 
In this case, the most likely scenario for the hard X-rays is 
a low-luminosity AGN (LLAGN hereafter) 
behind the dense gas near to the center. 
The absorption of the hard component is 
{\it N$_{\rm H}$}=$(2$--$3)\times 10^{22}$~{\rm cm$^{-2}$}. 
The {\it N$_{\rm H}$}\ value estimated from the CO observations 
at the position of the hard component (i.e. at the source~8) is 
{\it N$_{\rm H}$}=$(4$--$6)\times 10^{22}$~{\rm cm$^{-2}$}\ 
(Lo {\rm et al.}\  1987; Nakai {\rm et al.}\  1987). 
Assuming that about half of the {\it N$_{\rm H}$}\ is located 
in front of the M82 center, 
the ASCA result is consistent with the scenario of the absorption 
due solely to the molecular clouds. 

Several ASCA observations show that some spiral galaxies are hosts of LLAGN, 
whose luminosities are $10^{40}$ to $10^{41}$~{\rm erg~s$^{-1}$} 
(Ishisaki {\rm et al.}\  1996). 
ROSAT HRI detection of a short-time variability may also support 
this scenario (Collura {\rm et al.}\  1994). 
One may argue that no clear AGN activity has been reported in M82. 
However, the large absorption column and the fairly 
complicated starburst activities 
would make it hard to detect any LLAGN activity at other wavelengths. 

\subsection{The Soft and Medium Temperature Plasmas}
We found two-temperature extended plasmas from M82. 
Since the soft component is more extended than the medium component, 
the absorbing column for these components is 
larger than the galactic value of $4.3\times 10^{20}${\rm cm$^{-2}$}, 
it needs extra absorption near to M82 (Stark {\rm et al.}\  1992). 
The metal abundances of these hot plasmas were also constraint to some extent.
We will discuss these issue separately. 

\subsubsection{The X-ray images and relevant absorption}
We see that the counting rate in the 0.5--0.8~keV band 
at the peak of the other two band images in figure~6 is low. 
This would be due to the absorption of cool matter around the M82 center. 
Comparing the 0.5--0.8~keV band image with the CO and H~I images, 
we found that the low-brightness region in the 0.5--0.8~keV band
corresponds to the CO and H~I disk 
(Lo {\rm et al.}\  1987; Nakai {\rm et al.}\  1987; 
Weliachew {\rm et al.}\  1984). 
The {\it N$_{\rm H}$}\ value at the disk estimated from the CO and H~I 
is $(4$--$6)\times 10^{22}$~{\rm cm$^{-2}$}, 
which is large enough to absorb soft X-rays. 
Therefore, a simple idea is that the diffuse soft X-ray emitting plasma
has a flux peak at the center as well as a component of 
the medium-temperature plasma, 
but highly absorbed by molecular clouds around the center. 

The soft and medium components extend more 
than the thickness of the molecular disk of $\sim \pm 100$~pc. 
Its average column density between $|z|=100$--$500$~pc is 
{\it N$_{\rm H}$}$\sim 6\times 10^{21}$~{\rm cm$^{-2}$}\ 
(Nakai {\rm et al.}\  1987). 
Then, for the relevant X-rays, the mean absorption column 
by the molecular clouds would be 
{\it N$_{\rm H}$}$\sim 3\times 10^{21}$~{\rm cm$^{-2}$}, 
half of the total {\it N$_{\rm H}$}.
The ASCA observation shows the absorption due to the M82 cool gas 
to be {\it N$_{\rm H}$}$=(2.6^{+0.5}_{-0.4}) \times 10^{21}$~{\rm cm$^{-2}$}\ 
after subtracting the galactic absorption from the total absorption of 
$(3.0^{+0.5}_{-0.4}) \times 10^{21}$~{\rm cm$^{-2}$} , 
which agrees well with the results of CO observations. 

\subsubsection{Metal abundances}
It is expected that a large number of sequential supernovae, 
mostly type-IIs, explode as the result of starburst activity, 
synthesize a huge amount of metals and cause strong emission lines in X-rays. 
However, the observed metal abundances are all below the cosmic values, 
especially the abundance of oxygen, 
the most important product in type-II supernovae, 
is found to be extremely low. 
Suppose that all the energy of the hot plasma 
(i.e. the soft and medium components) is supplied by supernova explosions, 
and the origin of the plasma gas is a mixture of the supernova ejecta and
interstellar gas; then, the thermal energy of the hot plasma 
per one supernova explosion can be written as 
\begin{eqnarray}
\frac{3}{2}\frac{M_{\rm ej} + M_{\rm am}}{\mu m_{\rm H}} k_{\rm B}T 
	& \ ^{\displaystyle <}_{\displaystyle \sim}\  & E_{\rm SN} , 
\end{eqnarray}
where $M_{\rm ej}$ and  $M_{\rm am}$ are the masses of the ejecta and 
the mean mass of the ambient gas heated by a single supernova.

The ratio of the mass of each metal ($M^*$) 
to the total mass ( $M_{\rm total}$) is expressed as
\begin{eqnarray}
\frac{M^*}{M_{\rm total}} 
	& = & \frac{M^*_{\rm ej} + M^*_{\rm am}}{M_{\rm ej} + M_{\rm am}} 
	\ ^{\displaystyle >}_{\displaystyle \sim}\  
	\frac{M^*_{\rm ej}}{M_{\rm ej} + M_{\rm am}}, 
\end{eqnarray}
where $M^*_{\rm am}$ and $M^*_{\rm ej}$  are the mass of each metal 
in the ambient, and are synthesized by the supernova explosion.

From equations (2) and (3), we obtain
\begin{eqnarray}
\frac{M^*}{M_{\rm total}} 
	& \ ^{\displaystyle >}_{\displaystyle \sim}\  & \frac{3}{2}M^*_{\rm ej}
	\frac{1}{\mu m_{\rm H}}\frac{1}{E_{\rm SN}}k_{\rm B}T \nonumber \\
	& = & 4.9\times 10^{-3} 
	\left(\frac{E_{\rm SN}}{10^{51}{\rm erg}}\right)^{-1} \nonumber \\
	&   & \times \left(\frac{k_{\rm B}T}{1~{\rm keV}}\right)
	\left(\frac{M^*_{\rm ej}}{1~M{\odot}}\right).
\end{eqnarray}
We adopted the results of $m_u=50$~{\it M$_{\odot}$}\ 
by Tsujimoto {\rm et al.}\  (1995) as ${M^*_{\rm ej}}$, 
in which the synthesized masses are 
1.8, 0.23, 0.12, 0.12, 0.041, 0.0080, and 0.091~{\it M$_{\odot}$}\ 
for O, Ne, Mg, Si, S, Ar, and Fe, respectively. 
Using these values, equation 4 
and the cosmic abundances in Anders and Grevesse (1989), 
the lower limits on the metal abundances are estimated to be 
0.71, 0.51, 0.72, 0.66, 0.43, 0.31, and 0.18 for 
O, Ne, Mg, Si, S, Ar, and Fe, respectively, 
based on the assumption of $E_{\rm SN}=10^{51}$~erg 
and {\it k$_{\rm B}T$}$=1$~keV. 
The observed abundances of sulfur and argon agree with this assumption, 
but those of oxygen and iron are smaller 
than the results based on it. 

One may argue that the low abundances can be explained 
by the remaining dust in the hot gas.
The number of oxygens synthesized by type-II supernovae 
is about 10 times larger than those of magnesium, silicon, sulfur, and iron. 
The number density of oxygens observed in M82 is still higher than 
the sum of the number densities of magnesium, silicon, sulfur, and iron 
by factor of 1.6. 
Even in the extreme case that oxygen, magnesium, silicon, sulfur, and iron 
would be confined in the dust as an oxide, 
a large fraction of the oxygen could be free from the dust. 
Neon is also a free atom. 
Thus, the dust-condensation scenario can not explain 
the low metalicity of the oxygen and neon in M82. 

\subsubsection{Relative ratios of the metal abundances}
Reserving the potential problem of the absolute metal abundances, 
we further discuss the relative abundance. 
We compare the metal abundance ratios of M82 
and those of type-II supernova in figure~7. 
The metal abundances of type-II supernova are 
integrated and averaged over the possible SN mass range 
with a reasonable mass function (Tsujimoto {\rm et al.}\  1995). 
For a comparison, we also plot the values of type-Ia supernovae in figure~7, 
which are adopted from model W7 
in Nomoto {\rm et al.}\  (1984) and Thielemann {\rm et al.}\  (1986). 
From figure~3, we can see that the abundance of the M82 gas peaks 
at the medium elements, such as silicon and sulfur, 
which are found neither in type-II nor type-Ia supernova. 
This does not agree with the scheme 
that metal synthesise due to supernova explosions 
contributes to the metal abundances in M82. 

Since the values of the abundance ratios of light elements in M82 are 
between those of type-Ia and type-II supernova, 
one may argue that they can be explained 
by a mixture of the two types of supernova, although the type-Ia supernovae 
may be minor compared to type-II in a starburst galaxy. 
However, the abundance ratio of iron to silicon in M82 is well 
below both that of type-Ia and type-II supernova. 
Accordingly, we have no solution which satisfies
both the abundances of iron and medium elements with this scenario. 

We note that although other starburst galaxies show 
similar low metal abundances to that of M82 
(e.g. ASCA observation of NGC~253; Awaki {\rm et al.}\  1996), 
starbursts associated with the Seyfert activity 
(e.g. NGC~1068; Ueno {\rm et al.}\  1994) exhibit normal metal abundances. 
This may imply that starburst activities, or relevant metal abundance, 
may differ from object to object. 

One of candidates for the origin of metals 
in the intra-cluster medium (ICM hereafter) is 
ejecta from starburst activities during the early phase of galaxy evolution 
(e.g. Sarazin 1988 and references therein). 
Although M82 is a nearby galaxy and is not a proto-galaxy, 
it exhibits typical starburst activity and is a unique target 
to address this issue at this moment. 
The ASCA observations of four bright clusters of galaxies 
reveal that the average metal abundances of oxygen, neon, silicone, sulfur,  
and iron are 0.48, 0.62, 0.65, 0.25, and 0.32, respectively 
(Mushotzky {\rm et al.}\  1996). 
The observation of M82 shows lower abundances for all of the metals. 
In particular, oxygen and iron abundances of M82 are lower than 
those of ICM by a factor of 8--10. 
This suggests that the ejecta from starburst activity 
would not be the origin of ICM. 

From the large number of samples obtained with ASCA, 
Matsumoto {\rm et al.}\  (1997) showed that the metal abundances 
of the hot inter stellar medium (ISM hereafter) 
in elliptical galaxies are about 0.3. 
The average metal abundances of hot ISM in three bright elliptical galaxies 
obtained with ASCA are 0.37, 0.33, 0.41, and 0.40 
for oxygen, silicon, sulfur, and iron, respectively 
(Awaki {\rm et al.}\  1994). 
The abundances of silicon and sulfur in M82 are comparable 
to those of the elliptical galaxies, but those of oxygen and iron are less. 

\subsection{Non Equilibrium Ionization}
Since the NEI model predicts a higher electron temperature 
than the line-emitting ionization temperature, 
it may be conceivable 
that low-temperature NEI model can reproduce the overall spectra.
We thus tired to simultaneously fit the SIS~0+1 and GIS~2+3 spectra 
with a 2-temperature ionization non-equilibrium plasma model 
(2-NEI model in short). 
Since no NEI code based on the Raymond Smith thermal plasma model 
is available in the public XSPEC, 
we installed a NEI model based on the code of Masai (1984, 1994). 
We note that the total number of free parameters 
between 3-RS and 2-NEI are nearly the same as that of each other. 
This 2-NEI model gives a nice fit with $\chi^2$/d.o.f. = $441.3/390$, 
almost equal to the case of 3-RS. 
The best-fit results are given in table~4. 

The small size estimated from 
the observed time variability of a couple of years 
gives a severe constraint on the NEI model. 
The emission measure $n_{\rm e}^2V$ is $2\times 10^{63}$~{\rm cm$^{-3}$}, 
where $n_{\rm e}$ and $V$ are the density and volume of the plasma. 
If the size of the emitting region is as small as 1~pc, 
then $n_{\rm e}$ is larger than $5\times 10^3$~{\rm cm$^{-3}$}. 
Therefore, the observed $n_{\rm e}t$ value of 
$9\times 10^9$~{\rm s~cm$^{-3}$}\ constrains 
the age of the plasma to be smaller than three weeks.
This extremely short time scale led us to conclude 
that the NEI plasma is unrealistic.
Thus, we safely conclude that the origin of the hard X-rays is 
not a NEI plasma, but either a CIE plasma or a non-thermal source. 

\section{Conclusion}
The X-ray structure of M82 composes of a point-like, 
heavily absorbed hard X-ray source 
with a {\it k$_{\rm B}T$}=14~keV temperature 
or power-law with photon index of 1.7, 
and extended two-temperature plasmas 
with {\it k$_{\rm B}T$}=0.3~keV and 0.95~keV. 

We found a long-term variability and spectral change in the hard X-rays, 
and inferred that the most likely origin is 
due to an obscured low-luminosity AGN. 

The two-temperature plasmas would be produced 
by the starburst activity of M82. 
However, the observed low metal abundances in M82 gas 
are not easily explained by any mixture of supernova explosions. 

A two-temperature NEI model can reproduce the observed spectrum, 
but is physically unrealistic. 

\vspace{1pc}

The authors would like to thank Drs. R.~Petre, R.~Mushotzky, 
T.~J.~Ponman, R.~J.~Terlevich, D.~K.~Strickland, K.~Makishima, 
and T.~Ohashi for useful suggestions and discussions. 
We also thank Prof.~K.~Masai, who kindly supplied his NEI code to us. 
We wish to thank the ASCA software team, 
who developed the analysis software which we used. 
We are grateful to Prof.~Tanaka and the ASCA hardware team, 
who developed ASCA and supported its operation. 

\clearpage
\section*{References}
\small
\re Anders E., Grevesse N. 1989, 
	Geochimi. et Cosmochim. Acta 53, 197
\re Awaki H., Mushotzky R., Tsuru T., Fabian A.C., Fukazawa T., 
	Loewenstein M., Makishima K., Matsumoto H. et al. 1994, PASJ 46, L65 
\re Awaki H., Tsuru T., Koyama K., and ASCA team 1996, 
	in UV and X-Ray Spectroscopy of Astrophysical and Laboratory Plasmas, 
	ed. K.~Yamashita, and T.~Watanabe p327
\re Bregman J.N., Schulman E., Tomisaka K. 1995, ApJ 439, 155 
\re Burke B.E., Mountain R.W., Harrison D.C., Bautz M.W., Doty J.P., 
	Ricker G.R., Daniels P.J. 1991. IEEE Trans ED-38, 1069 
\re Collura A., Reale F., Schulman E., Bregman J.N. 1994, ApJ, 420, L63 
\re Fabbiano G. 1988, ApJ 330, 672 
\re Ishisaki Y., Makishima K., Iyomoto N., Hayashida K., Inoue H., 
	Mitsuda K., Tanaka Y., Uno S. et al. 1996, PASJ 48, 237
\re Kohmura Y., Inoue H., Aoki T., Ishida M., Kotani T., Tanaka Y., 
	Ishidaki Y., Makishima K., Matsushita K. 1994, PASJ 46, L157 
\re Kronberg P.P., Biermann P., Schwab F.R. 1985, ApJ 291, 693 
\re Kronberg P.P., Sramek R.A. 1992, 
	in X-Ray emission from Active Galaxtic Nuclei 
	and the Cosmic X-ray background, 
	ed. W.~Brinkman, J.~Tr\"umper MPE Conf.~235 p1992 
\re Lo K.Y., Cheung K.W., Masson C.R., Phillips T.G., Scott S.L., Woody D.P.
	1987, ApJ 312, 574 
\re Masai K. 1984, Ap\&SS 98, 367 
\re Masai K. 1994, ApJ 437, 770 
\re Makishima K., Tashiro M., Ebisawa K., Ezawa H., Fukazawa Y., Gunji S., 
	Hirayama M., Idesawa E. {\rm et al.}\  1996, PASJ 48, 171 
\re Matsumoto H., Koyama K., Awaki H., Tsuru T., 
	Loewenstein M., Matsushita K. 1997, ApJ 482, 133 
\re Moran E.C., Lehnert M.D. 1997, ApJ 478, 172 
\re Morrison R., MaCammon D. 1983, ApJ 270, 119 
\re Mushotzky R., Loewenstein M., Arnaud K.A., Tamura T., Fukazawa Y.,
        Matsushita K., Kikuchi K., Hatsukade I. 1996, ApJ 466, 686 
\re Muxlow T.W.B., Pedlar A., Wilkinson P.N., Axon D.J., Sanders E.M.,
        de~Bruyn A.G. 1994, MNRAS 266, 455 
\re Nakai N., Hayashi M., Handa T., Sofue Y., Hasegawa T. and Sasaki, M. 
	1987 PASJ 39, 685 
\re Nomoto K., Thielemann F.-K., Yokoi K. 1984, ApJ 286, 644 
\re Ohashi T., Ebisawa K., Fukazawa Y., Hiyoshi K., Horii M., 
	Ikebe Y., Ikeda H., Inoue H. {\rm et al.}\  1996, PASJ 48, 157 
\re O'Connell R.W., Gallagher III J.S., Hunter D.A., Colley W.N.
	1995, ApJ 446, L1 
\re Okada K., Mihara T., Makishima K., the ASCA team 1994, 
	in New Horizon of X-Ray Astronomy ed. F.~Makino, T.~Ohashi 
	(Universal Academy Press, Tokyo), p515 
\re Ptak A., Serlemitsos P., Yaqoob T., Mushotzky R., Tsuru T. 
	1997, AJ 113, 1286
\re Petre R. 1993, in The Nearest Active Galaxies 
	ed. H.~Netzer, J.~Beckman (CSIC, Madrid) p117
\re Petre R., Okada K., Mihara T., Makishima K., Colbert E.J.M. 
	1994, PASJ 46, L115 
\re Raymond J.C., Smith B.W. 1977, ApJS 35, 419 
\re Rieke G.H., Lebofsky M.J., Thompson R.I., Low F.J., Tokunaga A.T.
	1980, ApJ 238, 24 
\re Sarazin C.L. 1988, in X-ray emission from clusters of galaxies 
	(Cambridge University Press, New York) p223 
\re Serlemitsos P.J., Jalota L., Soong Y., Kunieda H., 
	Tawara Y., Tsusaka Y., Suzuki H., Sakima Y., Yamazaki T., 
	Yoshioka H. 1995, PASJ 47, 105 
\re Schaaf R., Pietsch W., Biermann P.L., Kronberg P.P., Schmutzler T.
	1989, ApJ 336, 722 
\re Strickland D.K., Ponman T.J., Stevens I.R. 1997, A\&A 320, 378. 
\re Stark A.A., Gammie C.F., Wilson R.W., Bally J., Linke R., 
	Heiles C., Hurwitz M. 1992, ApJS 79, 77 
\re Tanaka Y., Inoue H., Holt S.S. 1994, PASJ 46, L37 
\re Terlevich R.J. 1994 in 
	Circumstellar Media in the Late Stages of Stellar Evolution 
	ed.  R.E.S. Clegg, I.R. Stevens, W.P.S. Meikle p153 
\re Thielemann F.-K., Nomoto K., Yokoi K. 1986, A\&A 158, 17 
\re Tsujimoto T., Nomoto K., Yoshii Y., Hashimoto M., 
	Yanagida S., Thielemann F.-K. 1995, MNRAS 277, 945 
\re Tsuru T. 1992, PhD thesis, University of Tokyo, ISAS RN, 528 
\re Tsuru T., Hayashi I., Awaki H., Koyama K., Fukazawa Y., 
	Ishisaki Y., Iwasawa K., Ohashi T. {\rm et al.}\  
	1994, in New Horizen of X-Ray Astronomy, 
	ed. F.~Makino, T.~Ohashi (University Academy Press, Tokyo) p529
\re Tsuru T., Hayashi I., Awaki H., Koyama K., Fukazawa Y., 
	Ishisaki Y., Iwasawa K., Ohashi T. {\rm et al.}\  1996, 
	in X-ray Imaging and Spectroscopy of Cosmic Hot Plasmas, 
	ed. F.~Makino, K.~Mitsuda (University Academy Press, Tokyo) p157 
\re Ueno S., Mushotzky R.F., Koyama K., Iwasawa K., Awaki H., Hayashi I. 
	1994, PASJ 46, L71 
\re Watson M.G., Stanger V., Griffiths R.E. 1984, ApJ 286, 144 
\re Watson M.G., Warwick R.S., Stewart G.C., Folgheraiter E. 1994, 
        in New Horizon of X-Ray Astronomy 
	ed. F.~Makino, T.~Ohashi (University Academy Press, Tokyo) p533
\re Weliachew L., Fomalont E.B., Greisen E.W. 1984, A\&A 137, 225 


\clearpage

\section*{Figure Captions}
\small

Fig.~1 SIS~0+1 and GIS~2+3 spectra obtained within a $6'.0$ radius 
	from the center of M82 and the best-fit 3-RS model. \\


Fig.~2 Two-dimensional parameter map of $n_{\rm e}t$ 
	and {\it k$_{\rm B}T$}\ in the case of the NEI model. \\


Fig.~3 Metal abundances from the 3-RS model fit. \\


Fig.~4 Confidence contour maps between the iron K-emission 
	line center energy and its equivalent width 
	obtained with Ginga and ASCA. \\


Fig.~5 Three bands images of SIS~0 smoothed with $\sigma=0'.212$. 
The contour levels are 
0.00969, 0.0145, 0.0218, 0.0327, 0.0491, 0.0736, 0.110, 
0.166, 0.248, 0.373 counts~arcmin$^{-2}$~s$^{-1}$. 
The dashed lines show gaps among the CCD chips in SIS~0. \\


Fig.~6  Closed version of the 0.5--0.8~keV image of SIS~0 
smoothed with $\sigma=0'.159$. 
The contour levels are 
0.0194, 0.0242, 0.0291, 0.0339, 0.0388, 0.0436, 0.0485, 0.0522, 
0.0630, 0.0727, 0.0921 counts~arcmin$^{-2}$~s$^{-1}$. 
The dashed lines show the gaps among the CCD chips in SIS~0. 
The crossed mark indicates the peak position of the medium and hard 
components. \\


Fig.~7 Relative metal abundaces synthesized by 
	type-Ia and type-II supernovae. 
	The data have been adopted from Tsujimoto {\rm et al.}\  (1995), 
	Nomoto {\rm et al.}\  (1984), and Thielemann {\rm et al.}\  (1986). \\





\clearpage
\normalsize

\begin{table}[h]
 \begin{center}
	Table 1 The line energies and fluxes$^*$. \\
 \begin{tabular}{|c|c||c|}
\hline
Line Energy		& photon flux			& ID		\\
(keV)			& ($10^{-5}$ photon s$^{-1}$ cm$^{-2}$)
							&		\\
\hline
$1.351\pm 0.011$	& $7.5\pm 2.3$		& Mg K$\alpha$ (He-like) \\
$1.457\pm 0.043$	& $1.3^{+2.0}_{-1.3}$	& Mg K$\alpha$ (H-like) \\
$1.864\pm 0.008$	& $8.1^{+2.3}_{-2.4}$	& Si K$\alpha$ (He-like) \\
$2.006\pm 0.012$	& $5.8^{+1.9}_{-2.0}$	& Si K$\alpha$ (H-like) \\
$2.457\pm 0.016$	& $2.5^{+2.0}_{-1.9}$	& S  K$\alpha$ (He-like) \\
\hline
\end{tabular}
\end{center}
* All the errors including upper limits are described 
at 90\% confidence limits.\\
\end{table}

\begin{table}[h]
\begin{center}
	Table 2 The three Raymond-Smith model$^*$. \\
\begin{tabular}{|cc|ccc|}
\hline
\multicolumn{2}{|c|}{Parameter}
			&	\multicolumn{3}{c|}{Component}	\\
	&
			&	Soft	
			&	Medium	
			&	Hard		\\
\hline
\multicolumn{2}{|c|}{{\it L$_{\rm X}$} $^{\rm a}$ ($10^{40}$ {\rm erg~s$^{-1}$})}
			&	$1.6$
			&	$1.4$
			&	$3.6$	\\
\multicolumn{2}{|c|}{{\it N$_{\rm H}$}$_{\rm whole}$ ($10^{21}$ {\rm cm$^{-2}$})}
			&	
\multicolumn{3}{c|}{${3.0^{+0.5}_{-0.4}}^{\rm b}$} \\
\multicolumn{2}{|c|}{{\it N$_{\rm H}$}$_{\rm hard}$ ($10^{21}$ {\rm cm$^{-2}$})}
			&	\multicolumn{2}{c}{}
			&	$18.9^{+4.2}_{-4.2}$	\\
\multicolumn{2}{|c|}{{\it k$_{\rm B}T$} (keV)}
			&	$0.32^{+0.03}_{-0.03}$
			&	$0.95^{+0.05}_{-0.08}$
			&	$13.9^{+3.9}_{-2.6}$	\\
\hline
Metalicity (Cosmic)$^{\rm c}$	
& N  & \multicolumn{2}{c|}{$<0.15$} & \\
& O  & \multicolumn{2}{c|}{$0.063^{+0.033}_{-0.019}$} & \\
& Ne & \multicolumn{2}{c|}{$0.15^{+0.04}_{-0.03}$} & \\
& Mg & \multicolumn{2}{c|}{$0.25^{+0.07}_{-0.05}$} & $0.15^{\rm d}$\\
& Si & \multicolumn{2}{c|}{$0.40^{+0.07}_{-0.05}$} & \\
& S  & \multicolumn{2}{c|}{$0.47^{+0.12}_{-0.11}$} & \\
& Ar & \multicolumn{2}{c|}{$<0.51$} & \\
& Fe & \multicolumn{2}{c|}{$0.049^{+0.014}_{-0.009}$} &\\
\hline
\multicolumn{2}{|c|}{$\chi^2/{\rm d.o.f.}$}
			& \multicolumn{3}{c|}{$444.8/372$}	\\
\hline
\end{tabular}
\end{center}
* All the errors including upper limits are described 
at 90\% confidence limits.\\
a: Unabsorbed luminosity in the 0.5--10~keV band. \\
b: Galactic absorption of $4.3 \times 10^{20}$cm$^{-2}$ is included 
(Stark {\rm et al.}\  1992).
This parameter is common among the three components. \\
c: Relative abundance among the elements are fixed to be cosmic values. \\
d: Relative metal abundance ratios in the hard components is 
fixed to the cosmic values. 
\end{table}

\clearpage
\begin{table}[h]
\begin{center}
	Table 3 The two Raymond-Smith and power-law model$^*$. \\
\begin{tabular}{|cc|ccc|}
\hline
\multicolumn{2}{|c|}{Parameter}
			&	\multicolumn{3}{c|}{Component}	\\
	&		&	Soft	
			&	Medium	
			&	Hard		\\
\hline
\multicolumn{2}{|c|}{{\it L$_{\rm X}$} $^{\rm a}$ ($10^{40}$ {\rm erg~s$^{-1}$})}
			&	$1.6$
			&	$1.4$
			&	$4.2$	\\
\multicolumn{2}{|c|}{{\it N$_{\rm H}$}$_{\rm whole}$ ($10^{21}$ {\rm cm$^{-2}$})}
			&	
\multicolumn{3}{c|}{${3.0^{+0.5}_{-0.3}}^{\rm b}$} \\
\multicolumn{2}{|c|}{{\it N$_{\rm H}$}$_{\rm hard}$ ($10^{21}$ {\rm cm$^{-2}$})}
			&	\multicolumn{2}{c}{}
			&	$23.1^{+5.0}_{-4.5}$	\\
\multicolumn{2}{|c|}{{\it k$_{\rm B}T$} (keV) or $\Gamma$}
			&	$0.31^{+0.04}_{-0.02}$
			&	$0.95^{+0.05}_{-0.08}$
			&	$1.70\pm 0.09$	\\
\hline
Metalicity (Cosmic)	
&	N	&	\multicolumn{2}{c|}{$<0.15$} & \\
&	O	&	\multicolumn{2}{c|}{$0.061^{+0.035}_{-0.019}$} &\\
&	Ne	&	\multicolumn{2}{c|}{$0.14^{+0.05}_{-0.03}$} & \\
&	Mg	&	\multicolumn{2}{c|}{$0.25^{+0.07}_{-0.05}$} & \\
&	Si	&	\multicolumn{2}{c|}{$0.40^{+0.07}_{-0.06}$} & \\
&	S	&	\multicolumn{2}{c|}{$0.45^{+0.11}_{-0.10}$} & \\
&	Ar	&	\multicolumn{2}{c|}{$<0.48$} & \\
&	Fe	&	\multicolumn{2}{c|}{$0.048^{+0.014}_{-0.009}$} & \\
\hline
\multicolumn{2}{|c|}{$\chi^2/{\rm d.o.f.}$}
			& \multicolumn{3}{c|}{$454.2/373$}	\\
\hline
\end{tabular}
\end{center}
* All the errors including upper limits are described 
at 90\% confidence limits. \\
a: Unabsorbed luminosity in the 0.5--10~keV band. \\
b: Galactic absorption of $4.3 \times 10^{20}$cm$^{-2}$ is included 
(Stark {\rm et al.}\  1992).
This parameter is common among the three components. \\
\end{table}

\begin{table}[h]
\begin{center}
	Table 4 The two non-equilibrium ionization model$^*$. \\
\begin{tabular}{|cc|cc|}
\hline
\multicolumn{2}{|c|}{Parameter}
			&	\multicolumn{2}{c|}{Component}	\\
	&
			&	Soft	
			&	Hard		\\
\hline
\multicolumn{2}{|c|}{{\it L$_{\rm X}$} $^{\rm a}$ ($10^{40}$ {\rm erg~s$^{-1}$})}
			&	$1.7$
			&	$10.0$	\\
\multicolumn{2}{|c|}{{\it N$_{\rm H}$}$_{\rm whole}$ ($10^{21}$ {\rm cm$^{-2}$})}
			&	
\multicolumn{2}{c|}{${1.9\pm 0.2}^{\rm b}$} \\
\multicolumn{2}{|c|}{{\it N$_{\rm H}$}$_{\rm hard}$ ($10^{21}$ {\rm cm$^{-2}$})}
			&
			&	$14.8^{+1.9}_{-2.7}$	\\
\multicolumn{2}{|c|}{{\it k$_{\rm B}T$} (keV)}
			&	$0.65^{+0.03}_{-0.06}$
			&	$13.9^{+3.0}_{-1.5}$	\\
\multicolumn{2}{|c|}{${\rm log}(nt)$ ({\rm s~cm$^{-3}$})}
			&	$>11.6$
			&	$9.96^{+0.07}_{-0.11}$	\\
\hline
Metalicity (Cosmic)	
	&	N	&	\multicolumn{2}{c|}{$<0.41$} \\
	&	O	&	\multicolumn{2}{c|}{$0.29^{+0.05}_{-0.06}$}\\
	&	Ne	&	\multicolumn{2}{c|}{$0.22^{+0.09}_{-0.07}$} \\
	&	Mg	&	\multicolumn{2}{c|}{$0.17^{+0.03}_{-0.03}$} \\
	&	Si	&	\multicolumn{2}{c|}{$0.19^{+0.04}_{-0.02}$} \\
	&	S	&	\multicolumn{2}{c|}{$0.25^{+0.05}_{-0.06}$} \\
	&	Ar	&	\multicolumn{2}{c|}{$<0.49$}\\
	&	Fe	&    \multicolumn{2}{c|}{$0.075^{+0.011}_{-0.021}$}\\
\hline
\multicolumn{2}{|c|}{$\chi^2/{\rm d.o.f.}$}
			& \multicolumn{2}{c|}{$441.3/390$}	\\
\hline
\end{tabular}
\end{center}
* All the errors including upper limits are described 
at 90\% confidence limits. \\
a: Unabsorbed luminosity in the 0.5--10~keV band. \\
b: Galactic absorption of $4.3 \times 10^{20}$cm$^{-2}$ is included 
(Stark {\rm et al.}\  1992).
This parameter is common between the two components. \\
\end{table}




\clearpage

\begin{center}
  \epsfile{file=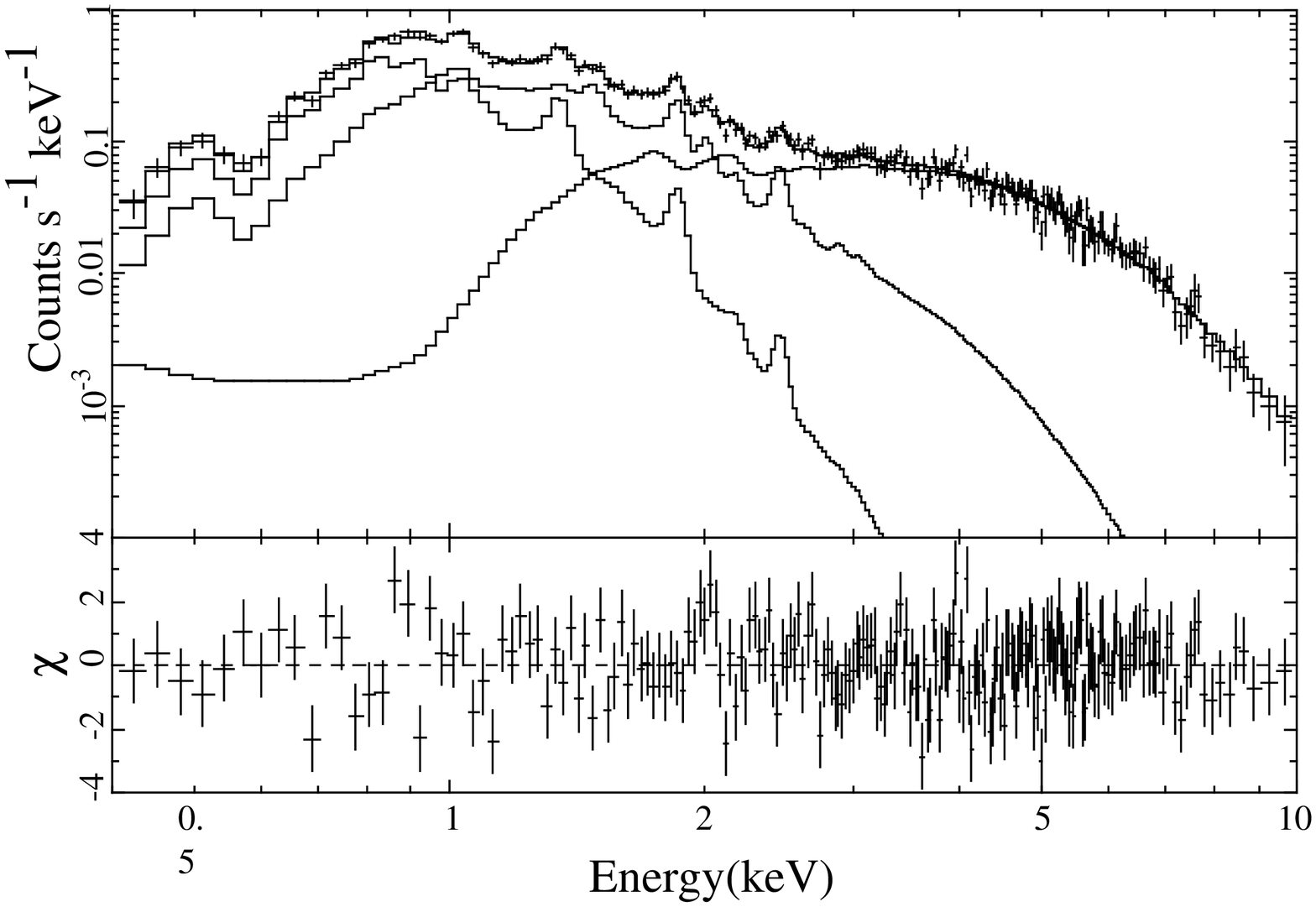,scale=0.3} 
  \epsfile{file=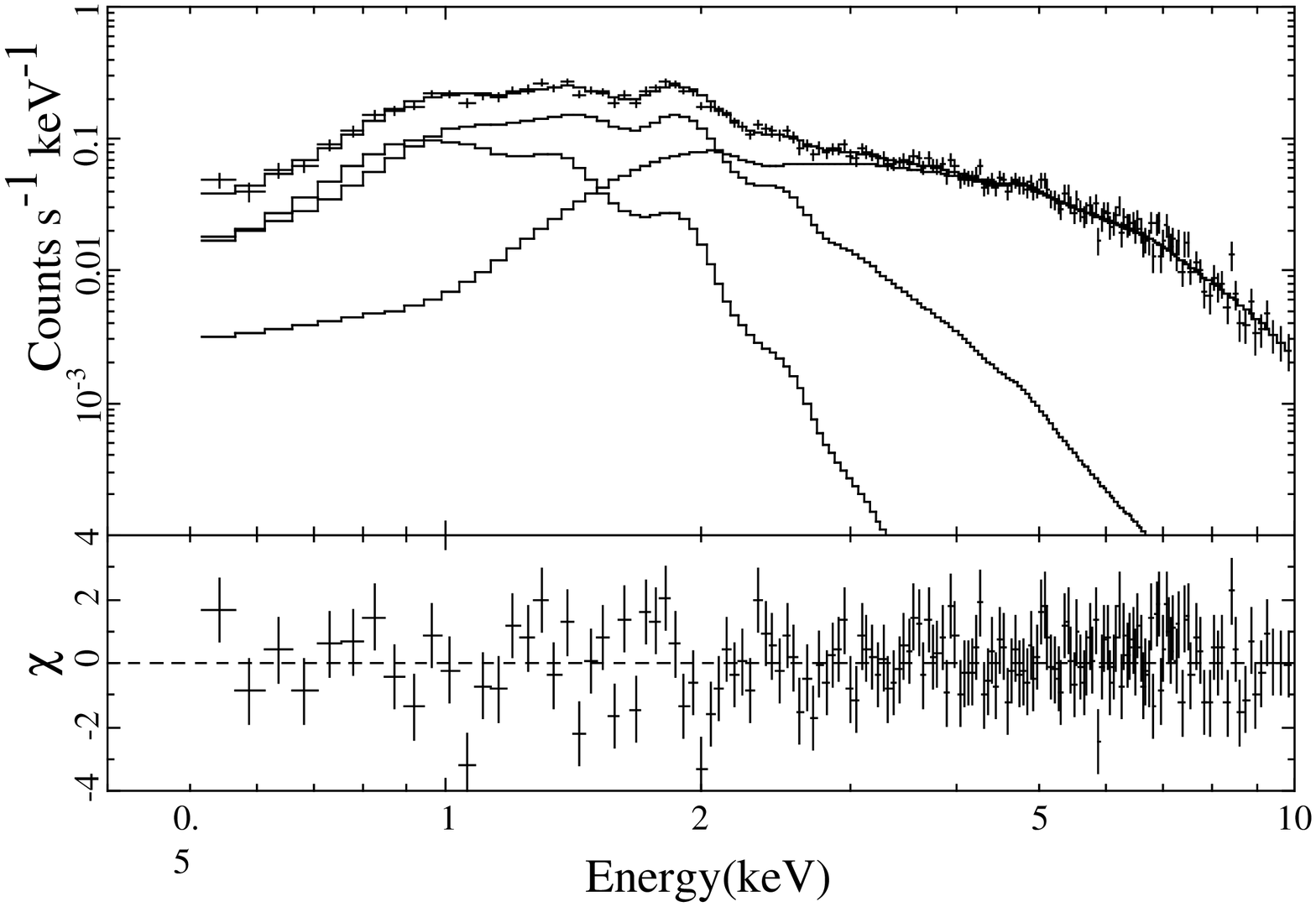,scale=0.3} \\
  Fugure~1: \\
\end{center}


\begin{center}
  \epsfile{file=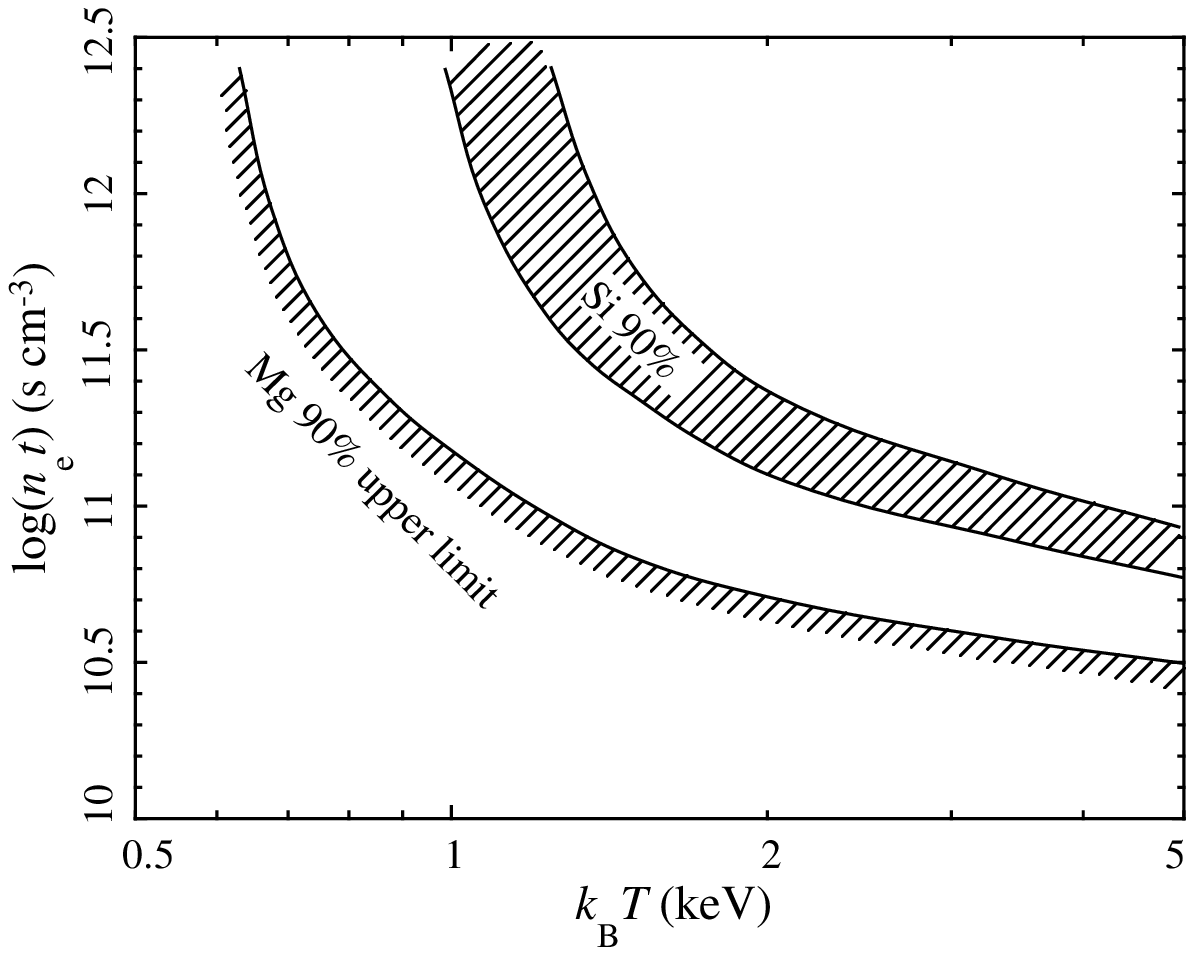,scale=0.7} \\
  Figure~2: \\
\end{center}


\begin{center}
  \epsfile{file=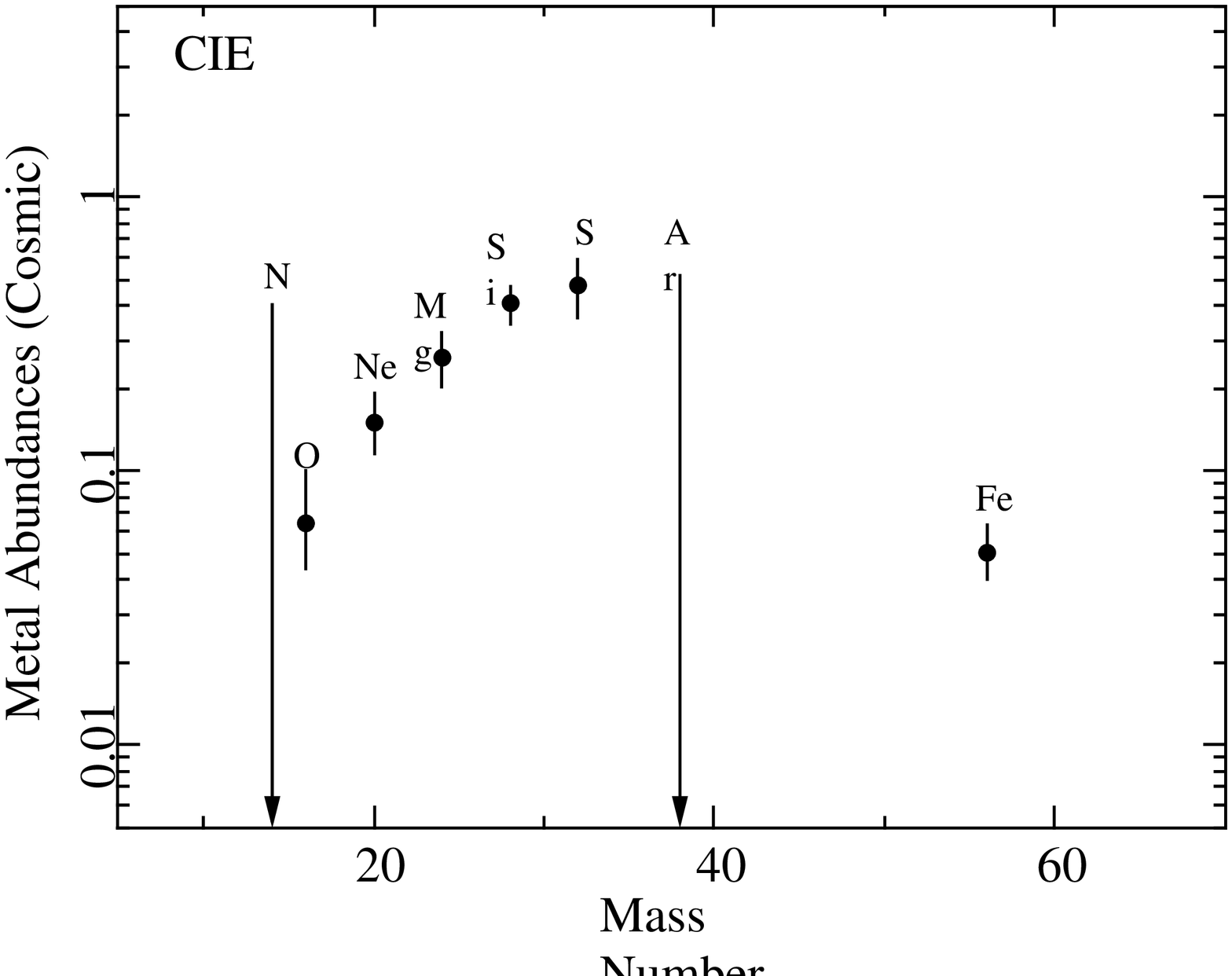,scale=0.4} \\
  Figure~3: \\
\end{center}


\begin{center}
  \epsfile{file=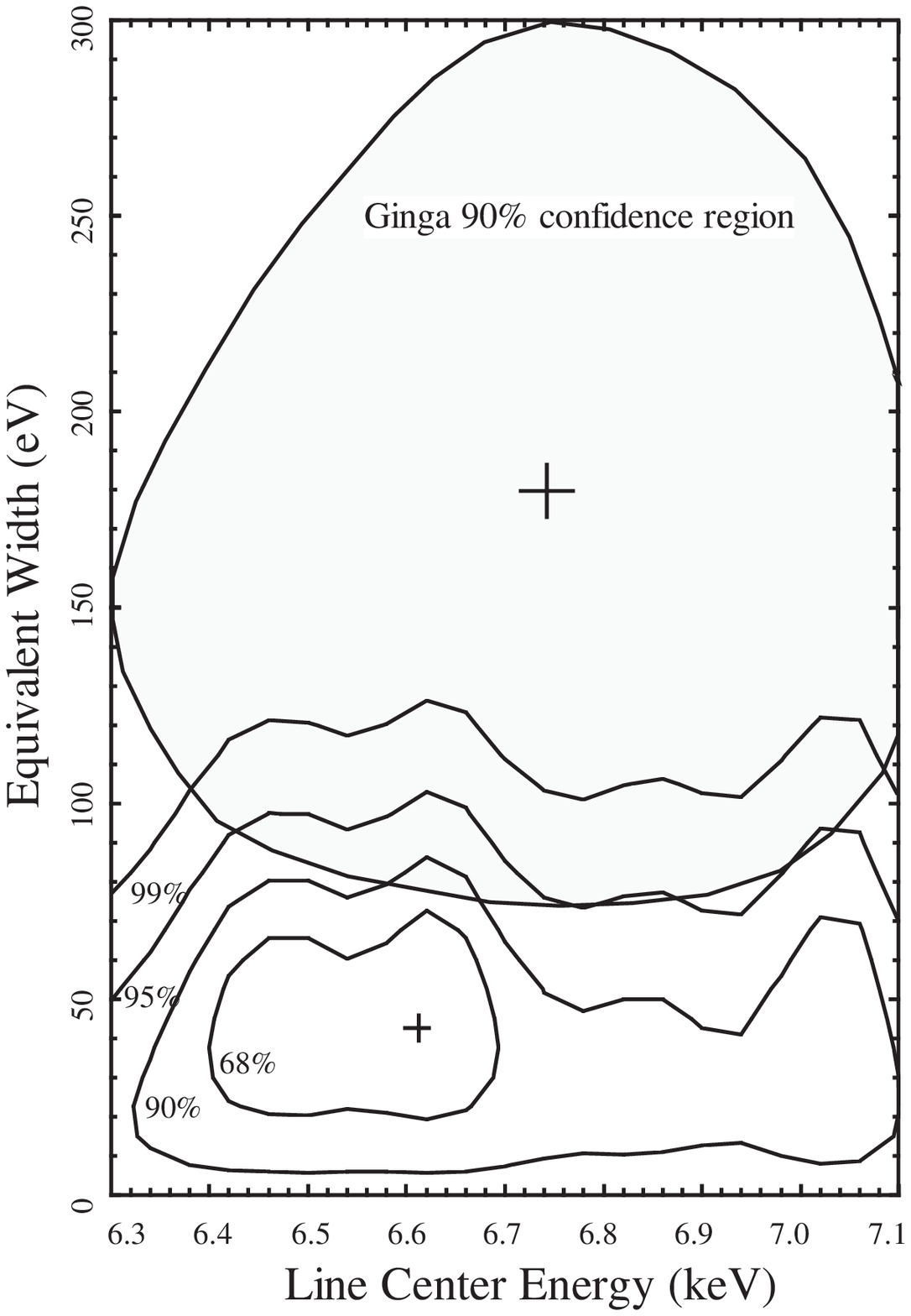,scale=0.4} \\
  Figure~4: \\
\end{center}


\begin{center}
  \epsfile{file=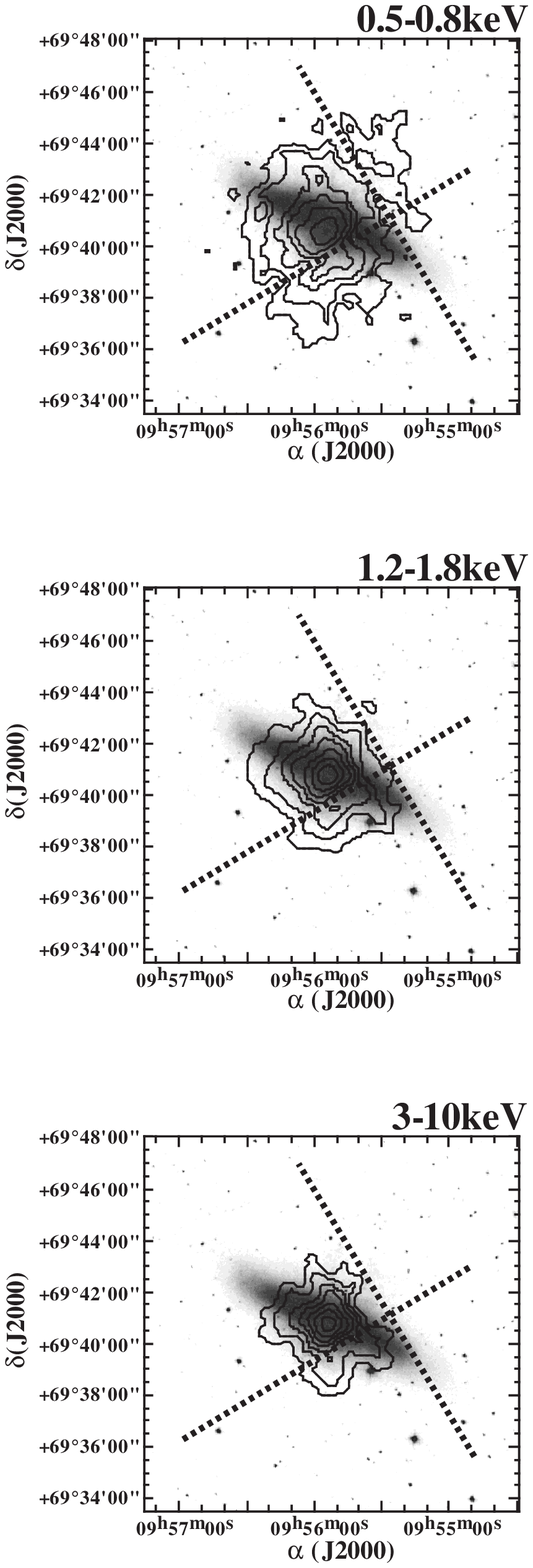,scale=0.7} \\
  Figure~5: \\
\end{center}


\begin{center}
  \epsfile{file=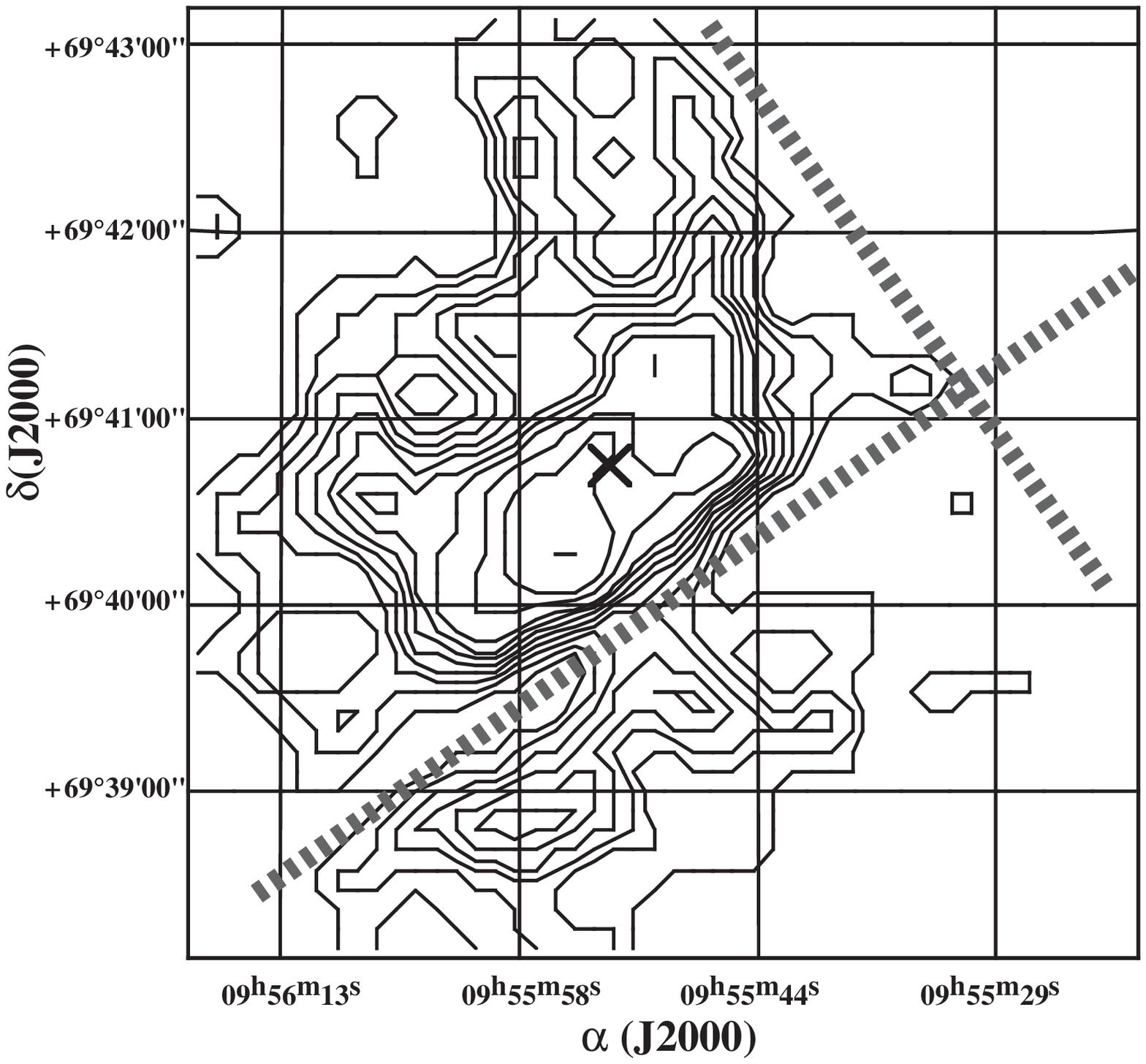,scale=0.5} \\
  Figure~6:  \\
\end{center}




\begin{center}
  \epsfile{file=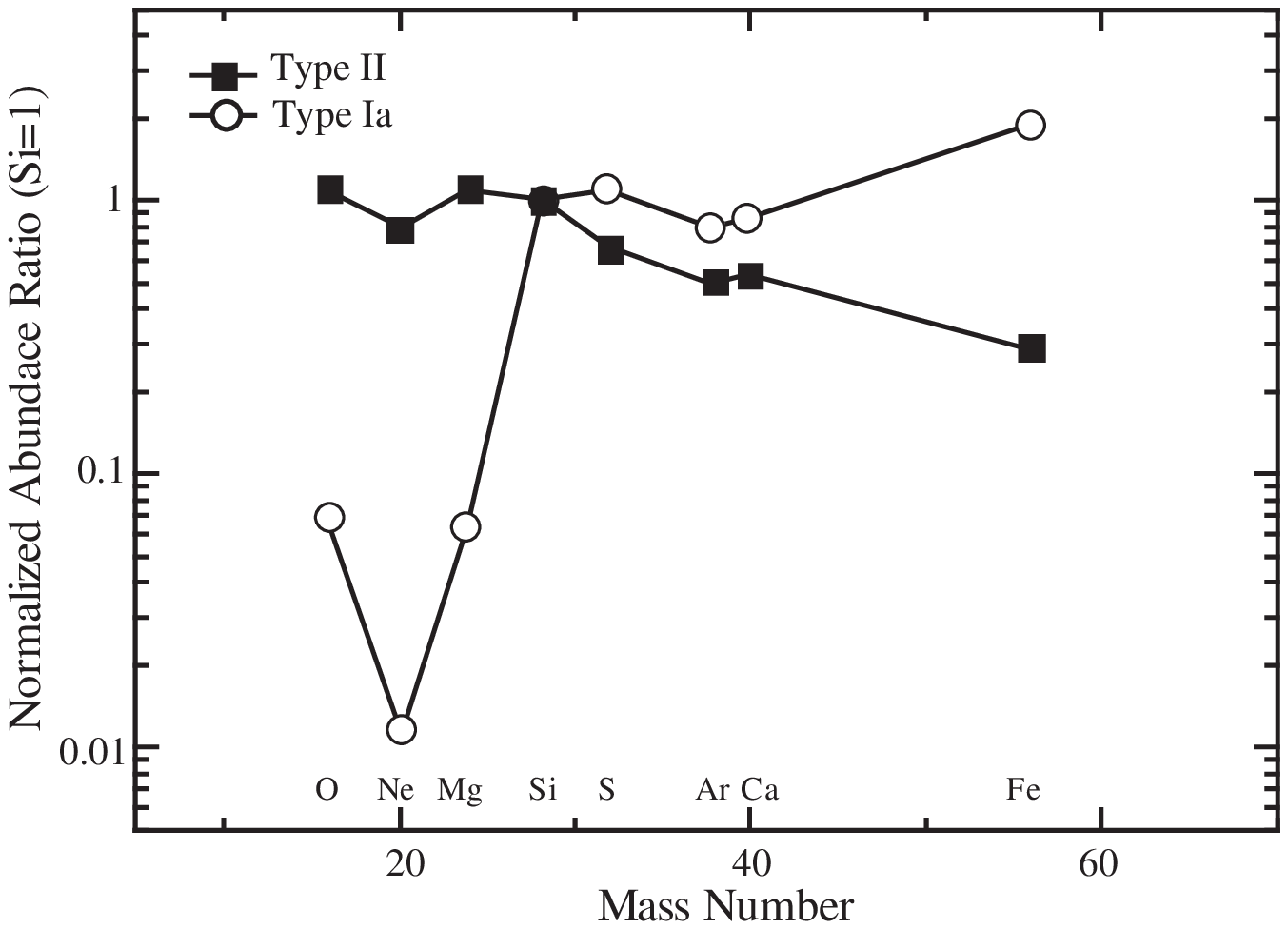,scale=0.7} \\
  Figure~7: \\
\end{center}


\end{document}